\documentclass[aps,amsmath,amssymb,reprint, double column,prl, showkeys]{revtex4-2}

\usepackage{graphicx,epsfig}
\usepackage{amssymb}
\usepackage{amsmath}
\usepackage{bm}
\usepackage{textcomp}
\usepackage{color}
\usepackage{float}

\usepackage{soul}

\begin{document}
\setcounter{page}{1}

\title[]{Composite Fermion Mass}
\author{K. A. \surname{Villegas Rosales}}
\author{P. T. \surname{Madathil}}
\author{Y. J. \surname{Chung}}
\author{L. N. \surname{Pfeiffer}}
\author{K. W. \surname{West}}
\author{K. W. \surname{Baldwin}}
\author{M. \surname{Shayegan}}
\affiliation{Department of Electrical and Computer Engineering, Princeton University, Princeton, New Jersey 08544, USA}

\date{\today}

\begin{abstract}

Composite fermions (CFs), exotic quasi-particles formed by pairing an electron and an even number of magnetic flux quanta emerge at high magnetic fields in an interacting electron system, and can explain phenomena such as the fractional quantum Hall state (FQHS) and other many-body phases. CFs possess an effective mass ($m_{CF}$) whose magnitude is inversely related to the most fundamental property of a FQHS, namely its energy gap. We present here experimental measurements of $m_{CF}$ in ultra-high quality two-dimensional electron systems confined to GaAs quantum wells of varying thickness. An advantage of measuring $m_{CF}$ over gap measurements is that mass values are insensitive to disorder and are therefore ideal for comparison with theoretical calculations, especially for high-order FQHS. Our data reveal that $m_{CF}$ increases with increasing well width, reflecting a decrease in the energy gap as the electron layer becomes thicker and the in-plane Coulomb energy softens. Comparing our measured masses with available theoretical results, we find significant quantitative discrepancies, highlighting that more rigorous and accurate calculations are needed to explain the experimental data. 

\end{abstract}

\maketitle  

Composite fermions (CFs) have captivated the attention of theorists and experimentalists in condensed matter physics for over three decades \cite{Jain.PRL.1989, HLR, Jain.CFbook.2007, Jain.CFchapter.2020,Shayegan.book.2020}. When a two-dimensional electron system (2DES) is subjected to a large perpendicular magnetic field ($B$), the ground states are typically dominated by electron-electron interaction. In the lowest orbital Landau level (LL), on the flanks of filling factor $\nu=1/2$, there are series of fractional quantum Hall states (FQHSs) at $\nu=p/(2p+1)$ where $p$ takes integer values \cite{Jain.PRL.1989, HLR, Jain.CFbook.2007, Jain.CFchapter.2020, Tsui.PRL.1982, Du.PRL.1993, Du.PRL.1993,Willett.PRL.1993, Kang.PRL.1993,Kamburov.PRL.2014,Jo.PRL.2018, Shayegan.book.2020,Goldman.PRL.1994,Smet.PRL.1999,Willett.PRL.1999,Kamburov.PRL.2014.b,Gokmen.NatPhys.2010,Du.PRL.1994,Manoharan.PRL.1994,Shafayat.NatPhys.2021,Deng.PRL.2016,Shafayat.PRL.2020,Du.SSC.1994,Leadley.PRL.1994,Maryenko.PRL.2012,Pan.PRL.2020,Kevin.PRL.2021,Chung.NatMat.2021, Coleridge.PRB.1995}. (The filling factor $\nu=nh/eB$ is defined as the number or fraction of the occupied LLs; $n$ is the 2DES density.) By attaching two flux quanta to each electron to form a CF, the FQHSs of electrons at electron filling factor $\nu$ can be explained as the integer QHSs of CFs at CF filling factor $p$ \cite{Jain.PRL.1989, HLR, Jain.CFbook.2007, Jain.CFchapter.2020}. Moreover, at and near $\nu=1/2$, the CFs have a well-defined Fermi sea \cite{HLR, Jain.CFbook.2007, Jain.CFchapter.2020} whose intriguing properties have been explored in numerous measurements \cite{Du.PRL.1993,Willett.PRL.1993, Kang.PRL.1993,Kamburov.PRL.2014,Jo.PRL.2018, Shayegan.book.2020,Goldman.PRL.1994,Smet.PRL.1999,Willett.PRL.1999,Kamburov.PRL.2014.b,Gokmen.NatPhys.2010,Du.PRL.1994,Manoharan.PRL.1994,Shafayat.NatPhys.2021,Deng.PRL.2016,Shafayat.PRL.2020,Du.SSC.1994,Leadley.PRL.1994,Maryenko.PRL.2012,Pan.PRL.2020,Kevin.PRL.2021,Chung.NatMat.2021, Coleridge.PRB.1995}.  Very recently, the experimentally-elusive Bloch ferromagnetism was demonstrated in CFs confined to a very high-quality, dilute GaAs 2DES \cite{Shafayat.NatPhys.2021}. Also, CFs have been used to probe the delicate periodic structure of a Wigner crystal hosted in a nearby 2DES \cite{Deng.PRL.2016}.

\begin{figure}[t!]
  \centering
    \psfig{file=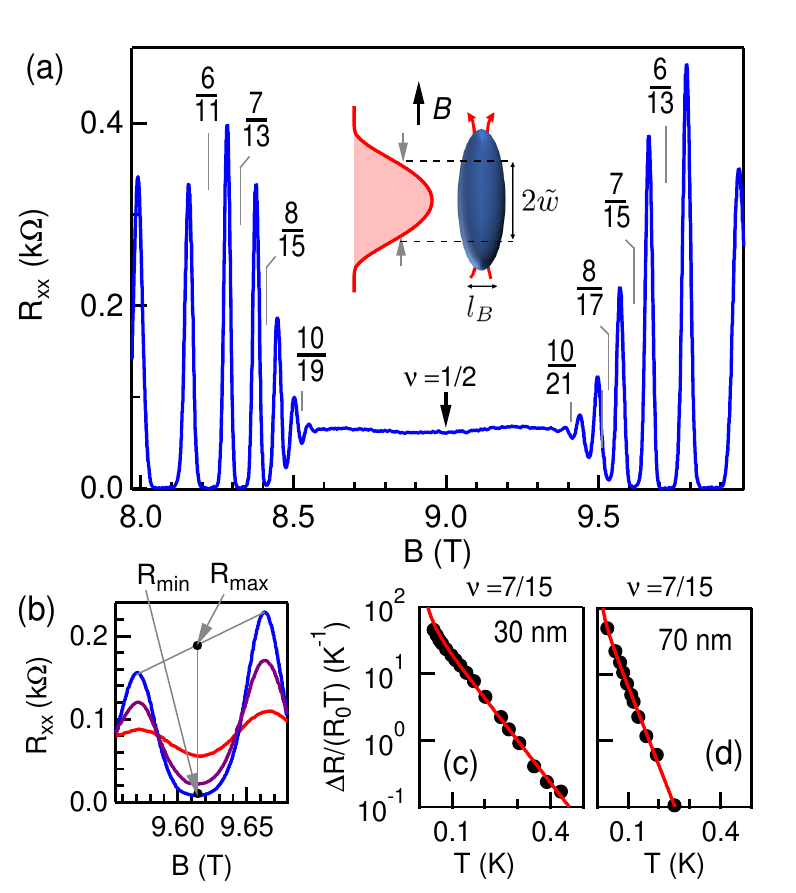, width=0.48\textwidth}
  \centering
  \caption{\label{transport} 
   (a) Longitudinal resistance $R_{xx}$ vs. \textit{B}, at $T\simeq25$ mK, for $S_{30}$, a GaAs 2DES with density $\simeq1.1\times10^{11}$ cm$^{-2}$ confined to a 30-nm-thick QW. Several minima in $R_{xx}$ are marked indicating FQHSs up to $\nu=10/19$ and $10/21$. The electron's charge distribution (from self-consistent calculations at $B=0$) for a $30$-nm-wide QW is shown as an inset. The electron layer thickness ($\tilde w$) is defined as the standard deviation of the charge distribution from its center. The cartoon to the right of the charge distribution depicts a two-flux CF in an electron layer with finite thickness. (b) $R_{xx}$ vs. $B$ traces near $\nu=7/15$ for $S_{30}$, at different temperatures: $T\simeq99$, $143$, and $248$ mK. The variables involved in the Dingle analysis, $R_{max}$ and $R_{min}$, are marked with black circles. (c,d) Plots of $\Delta R/(R_{0} T)$ vs. $T$, and the fits to the Dingle expression to extract $m_{CF}$ at $\nu=7/15$ for $S_{30}$ and $S_{70}$. The deduced $m_{CF}$ are $0.66$ and $1.22$ for $S_{30}$ and $S_{70}$.
  }
  \label{fig:transport}
\end{figure}

A very fundamental parameter characterizing CFs is their effective mass ($m_{CF}$), which is the focus of our work. This mass arises primarily from electron-electron interaction, and its magnitude determines the energy separation between the “CF LLs” (sometimes referred to as ``$\Lambda$ levels" \cite{Jain.CFbook.2007}), which in turn determines the size of the energy gaps for the FQHSs at $\nu=p/(2p+1)$. In realistic 2DESs with non-zero electron layer thickness ($\tilde w$) and finite separation between LL energies, $m_{CF}$ can be larger than its ideal value as both the softening of the in-plane Coulomb interaction and the mixing between the LLs can reduce the FQHS energy gaps and thus increase $m_{CF}$ \cite{Jain.CFbook.2007,Park.Activation.1999, Morf.PRB.2002}. Unlike the FQHSs' energy gaps which are believed to be further reduced by the small but ubiquitous sample disorder, $m_{CF}$ should be immune to small disorder \cite{Du.PRL.1994}. Despite this fundamental nature of $m_{CF}$, there have been no systematic measurements of its magnitude. We report here $m_{CF}$ measurements, via a Dingle analysis of the amplitude of the FQHS resistance oscillations, in extremely low-disorder samples as a function of $\tilde w$. By tuning $\tilde w$, the strength of the Coulomb interaction can be controlled. We show that indeed disorder does not seem to affect the magnitude of $m_{CF}$, in contrast to the energy gaps which are reduced by disorder. We then make a close comparison of the measured $m_{CF}$ with the results of available calculations \cite{Park.Activation.1999, Morf.PRB.2002}. We find that the measured $m_{CF}$ are typically larger than the calculated values, suggesting the need for more rigorous and accurate calculations to account for the experimental data.

We studied ultra-high-quality 2DESs confined to modulation-doped GaAs quantum wells (QWs), with well widths ($w$) ranging from $20$ to $70$ nm, grown on GaAs (001) substrates. The QWs are flanked by 150-nm-thick Al$_{0.24}$Ga$_{0.76}$As barriers, and the dopants are placed in doping wells \cite{Chung.PRM.2020}. The 2DESs have an electron density $n\simeq1.1\times10^{11}$ cm$^{-2}$ and transport mobility $\mu$ $\simeq6$ to $18\times10^{6}$ cm$^{2}$/Vs \cite{Kevin.PRL.2021}. We refer to samples with different $w$ by $S_{w}$. In the main text, we primarily present results from $S_{30}$. Data for some of the other samples are shown in the Supplementary Material (SM) \cite{SM}. The samples have a $4\times4$ mm$^{2}$ van der Pauw geometry, with alloyed InSn electrical contacts at the corners and edge midpoints. We used $^{3}$He and $^{3}$He-$^{4}$He dilution refrigerator systems, and conventional lock-in techniques for magnetoresistance measurements.


In Fig. 1(a) we present a longitudinal resistance ($R_{xx}$) vs. $B$ trace for $S_{30}$ near $\nu=1/2$ at $T\simeq25$ mK. The trace exhibits numerous minima corresponding to FQHSs at $\nu=p/(2p+1)$ as marked in Fig. 1(a). The parameter $p$ is the $\Lambda$-level filling factor for CFs \cite{Jain.PRL.1989,Jain.CFbook.2007,Jain.CFchapter.2020}. We find $R_{xx}$ minima up to $\nu=10/21$ and $10/19$ ($p=+10$ and $-10$), highlighting the very high quality of our 2DES. The sequence of high-order FQHSs, e.g., $\nu=10/21$ to $8/17$ and $10/19$ to $8/15$, appear as resistance oscillations emanating from $\nu=1/2$. Further away from $\nu=1/2$, well-developed FQHSs show vanishingly small $R_{xx}$. The trace in Fig. 1(a) has a striking resemblance to the Shubnikov-de Haas resistance oscillations and integer QHSs emerging from a Fermi sea (of CFs) around zero effective magnetic field, where $B_{eff}=B-B_{1/2}$ \cite{Du.PRL.1993}. These experimental signatures are in accordance with the well-established theory of CFs \cite{Jain.PRL.1989,Jain.CFbook.2007,Jain.CFchapter.2020,HLR} that treats the FQHSs of electrons as the integer QHSs of CFs. Within the CF picture, the energy gap for a given FQHS is the CF cyclotron energy $\hbar \omega_{CF} = \hbar e B_{eff}/m_{CF}$. 

We employ the standard procedure used to measure the effective mass of electrons near $B=0$ \cite{Dingle.1952, Shoenberg.1984} to quantitatively deduce $m_{CF}$ around $\nu=1/2$. This entails an analysis of the temperature dependence of the amplitude of resistance oscillations at specific fractional $\nu$ using the Dingle expression \cite{Dingle.1952, Shoenberg.1984}: $\Delta R/R_{0}=4 \times exp(-\pi/\omega_{CF}\tau _{q})\times \xi/sinh(\xi)$. The factor $\xi/sinh(\xi)$ represents the $T$-induced damping, where $\xi= 2\pi^2k_{B}T/\hbar\omega_{CF}$, and $\tau _{q}$ is the CF quantum lifetime. Other relevant parameters are defined as $\Delta R=(R_{max}-R_{min})$ and $R_{0}=(R_{max}+R_{min})/2$; see Fig. 1(b) for the definition of $R_{max}$ and $R_{min}$. The Dingle analysis of the FQHSs near $\nu=1/2$ has been successfully implemented in numerous studies \cite{Du.SSC.1994,Du.PRL.1994,Manoharan.PRL.1994,Leadley.PRL.1994,Coleridge.PRB.1995,Maryenko.PRL.2012,Coleridge.PRB.1995} to obtain $m_{CF}$.

Figures 1(b-d) present the Dingle analysis applied to the resistance oscillations around $\nu=7/15$ for $S_{30}$ and $S_{70}$. The trace in Fig. 1(b) shows the temperature dependence of $R_{xx}$. We plot $\Delta R/(R_{0}T)$ vs. $T$ on a semilog plot in Figs. 1(c) and (d), and fit the data to the Dingle expression to determine $m_{CF}$. The derived masses for $S_{30}$ and $S_{70}$ are $m_{CF}=0.66$ and $1.22$, in units of free electron mass $m_{0}$ which we will use throughout this paper. The mass is larger for the wider QW, as a direct consequence of the softening of the short-range Coulomb interaction for a thicker electron layer \cite{ZD}. Note that in our experiments the density is kept fixed while $w$ changes for different samples.

In Fig. 2(a), $m_{CF}$ measured from Dingle analysis are shown as open circles at many $\nu$ as a function of $B^{1/2}$ for $S_{30}$. Our measured $m_{CF}$ are about 10 to 20 times larger than the electrons' effective band mass ($m_{b}=0.067$) in GaAs. Despite some scatter, the $m_{CF}$ data points follow an approximately linear trend as a function of $B^{1/2}$ [dashed line in Fig. 2(a)] in accordance with the expected dependence of $m_{CF}$ on the Coulomb energy $E_{C}=e^2/4\pi\epsilon_{0}\epsilon l_{B}\propto B^{1/2}$, where $l_{B}=\sqrt{\hbar/eB}$ is the magnetic length and $\epsilon$ is the dielectric constant ($\epsilon=13$ for GaAs) \cite{HLR}. In the SM \cite{SM}, we report additional data for $S_{60}$ and $S_{70}$.


Before summarizing our measured $m_{CF}$ for different samples and comparing them to the results of the calculations, we would like to demonstrate that the measured $m_{CF}$ are not affected by disorder.  For an ideal 2DES, the energy gaps ($^{\nu}\Delta$) for the FQHSs are expected to scale as: $^{\nu}\Delta=(C/|2p+1|)$(\textit{e}$^{2}$/$4\pi\epsilon_{0}\epsilon$\textit{l}$_{B}$) \cite{HLR,Jain.CFbook.2007}, where $C\simeq0.3$ and $\nu=p/(2p+1)$. Figure 2(b) displays $^{\nu}\Delta$ vs. (\textit{e}$^{2}$/$4\pi\epsilon_{0}\epsilon$\textit{l}$_{B}$)/$(2p+1)$ (black symbols) measured for sample $S_{30}$ for $\nu$ up to $8/17$ and $8/15$ using two different techniques. The black triangles represent $^{\nu}\Delta$ determined from the standard procedure of making Arrhenius plots of $R_{xx}$ minimum as a function of temperature and fitting the data to $R_{xx}\propto exp(-^{\nu}\Delta/2T)$ \cite{Du.PRL.1993,Manoharan.PRL.1994,Pan.PRL.2020,Kevin.PRL.2021,Boebinger.PRL.1985,Willett.PRB.1988}. In Fig. 2(b), we also show black lines representing fits to the measured gaps. The magnitude of the negative intercepts of these lines with the $y$-axis provides an estimate of the phenomenological disorder parameter, $\Gamma$, based on the assumption that disorder reduces the gaps for different FQHSs by a fixed amount equal to $\Gamma$ \cite{Du.PRL.1993, Manoharan.PRL.1994, Kevin.PRL.2021, Pan.PRL.2020}. For the data of Fig. 2(b), we find $\Gamma=(0.7\pm0.2)$ K. 

In Fig. 2(b), using red open circles, we also plot values for $^{\nu}\Delta$ deduced from our measured $m_{CF}$ (Fig. 2(a)) and using the expression $^{\nu}\Delta=\hbar \omega_{CF} = \hbar e B_{eff}/m_{CF}$. The red lines, which are fits to the gaps deduced from $m_{CF}$, intercept the $y$-axis at $(0.04\pm0.13$) K, i.e., effectively at zero. (In the SM we present similar plots for $S_{60}$ and $S_{70}$ showing essentially the same behavior as in Fig. 2. The nearly-zero value of the intercept suggests that the measured $m_{CF}$ in our samples are insensitive to disorder. A similar conclusion was reached by Du \textit{et al.} \cite{Du.PRL.1994} from an analysis of the FQHSs' energy gaps and $m_{CF}$.

Figure 3(a) highlights the first main finding of our study: it displays the dependence of the measured $m_{CF}$ (red open circles) on electron layer thickness ($\tilde w$). We focus here on data at $\nu=3/7$ as a representative filling factor; data at other fillings are included in the SM \cite{SM}. We have chosen to focus on $\nu=3/7$ because: ($i$) it is relatively far from $\nu=1/2$ near which there is an apparent divergence of $m_{CF}$; ($ii$) the resistance oscillation surrounding $\nu=3/7$ is well behaved; and ($iii$) there is available theoretical data as we discussed below. Note that $\tilde w$ in Fig. 3 is given in units of the magnetic length $\l_B$. We use a Schroedinger-Poisson solver \cite{Schroedinguer.Poisson.solver} to calculate the charge distribution in a QW self-consistently at $B=0$, and define $\tilde w$ as the standard deviation of the charge distribution from its center. The charge distribution for $S_{30}$ is shown in Fig. 1(a) inset. As seen in Fig. 3(a), the measured $m_{CF}$ increases with increasing $\tilde w$, manifesting the weakening of the Coulomb interaction in samples with larger $\tilde w$. 

In Fig. 3(a), we have also included data points from three previous studies which reported $m_{CF}$ from Dingle analysis \cite{Du.SSC.1994, Leadley.PRL.1994, Coleridge.PRB.1995, footnote1}. These studies used 2DESs confined to GaAs/AlGaAs heterojunctions. To determine $\tilde w$ for the 2DESs in these samples, we used a Fang-Howard wavefunction \cite{ZD}. In Fig. 3(a) the data points from Refs. \cite{Du.SSC.1994, Leadley.PRL.1994, Coleridge.PRB.1995} are clearly consistent with our results. Given that the sample of Ref. \cite{Coleridge.PRB.1995} has about twice smaller mobility than our $S_{20}$ sample, the consistency seen in Fig. 3(a) provides additional evidence that disorder is playing a minimal role in affecting the measured $m_{CF}$. We would like to emphasize that the disorder independence of the effective mass deduced from the Dingle analysis has also been reported for \textit{electrons} (near $B=0$) in numerous studies \cite{Tan&Stormer.PRL.2005, Shashkin.PRB.2007, Melnikov.JETP.Lett.2014}. There is also theoretical justification \cite{Shoenberg.1984} that, at least for certain disorder broadening of the LLs, the two factors in the Dingle expression, $exp(-\pi/\omega_{CF}\tau _{q})$ and $\xi/sinh(\xi)$, are disentangled, so that a plot of the $T$-dependent damping of the magnetoresistance oscillations based on the $\xi/sinh(\xi)$ factor (see. e.g. Figs. 1(c, d)) would yield a mass that is independent of disorder.

\begin{figure}[t!]
  \centering
    \psfig{file=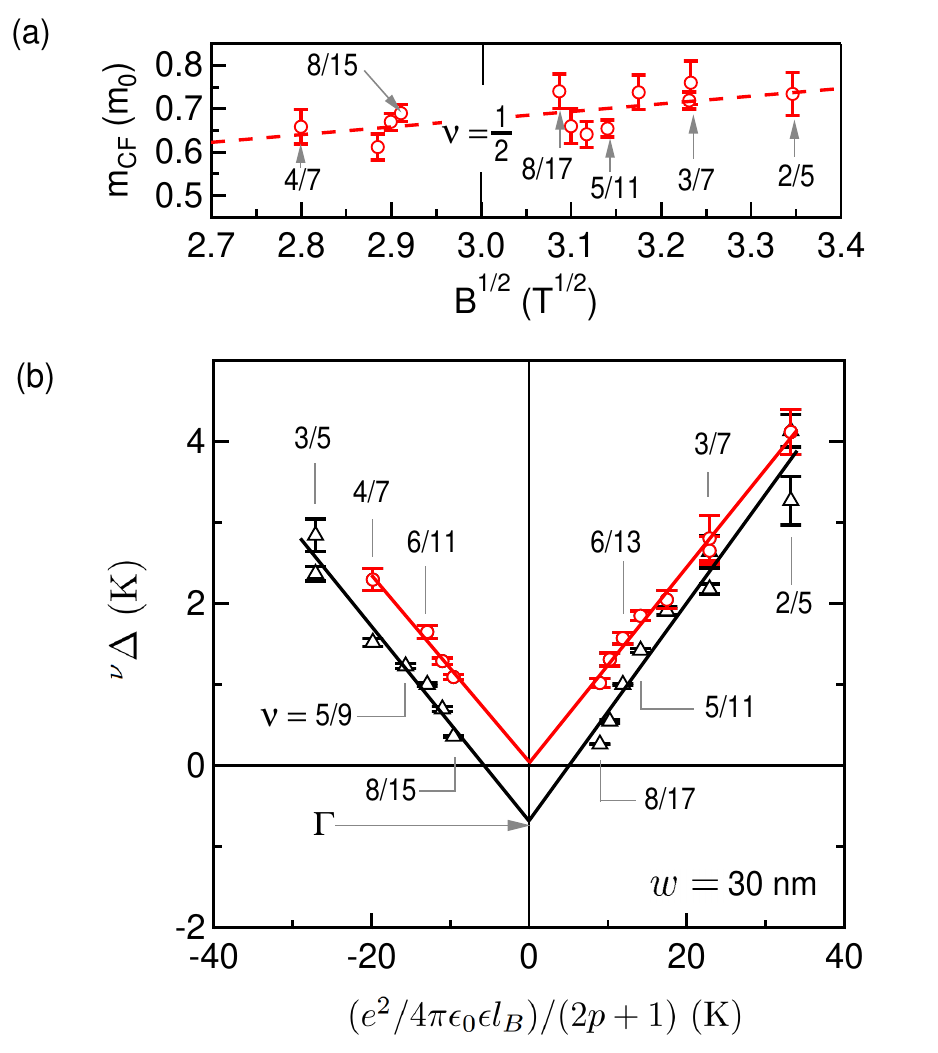, width=0.48\textwidth}
  \centering
  \caption{\label{Ip} 
(a) CF effective mass $m_{CF}$ vs. $B^{1/2}$ for a $S_{30}$. The open circles represent $m_{CF}$ determined from Dingle analysis at FQHS fillings. The dashed line is a fit through the data points. (b) Black triangles are the energy gaps $^{\nu}\Delta$, measured from Arrhenius plots of $R_{xx}$ vs. $1/T$ \cite{Kevin.PRL.2021}. They are shown as a function of $(e^{2}/4\pi\epsilon_{0}\epsilon l_{B})/(2p+1)$, where $\nu=p/(2p+1)$. The black lines are linear fits to the black data points. The open red circles are $^{\nu}\Delta$ deduced from the cyclotron energy of CFs, $\hbar \omega_{CF}=\hbar e B_{eff}/m_{CF}$, where we use the values of $m_{CF}$ shown in (a). The red lines are linear fits to the open circles.
  }
  \label{fig:Ip}
\end{figure}

\begin{figure*}[phbtp!]
  \centering
    \psfig{file=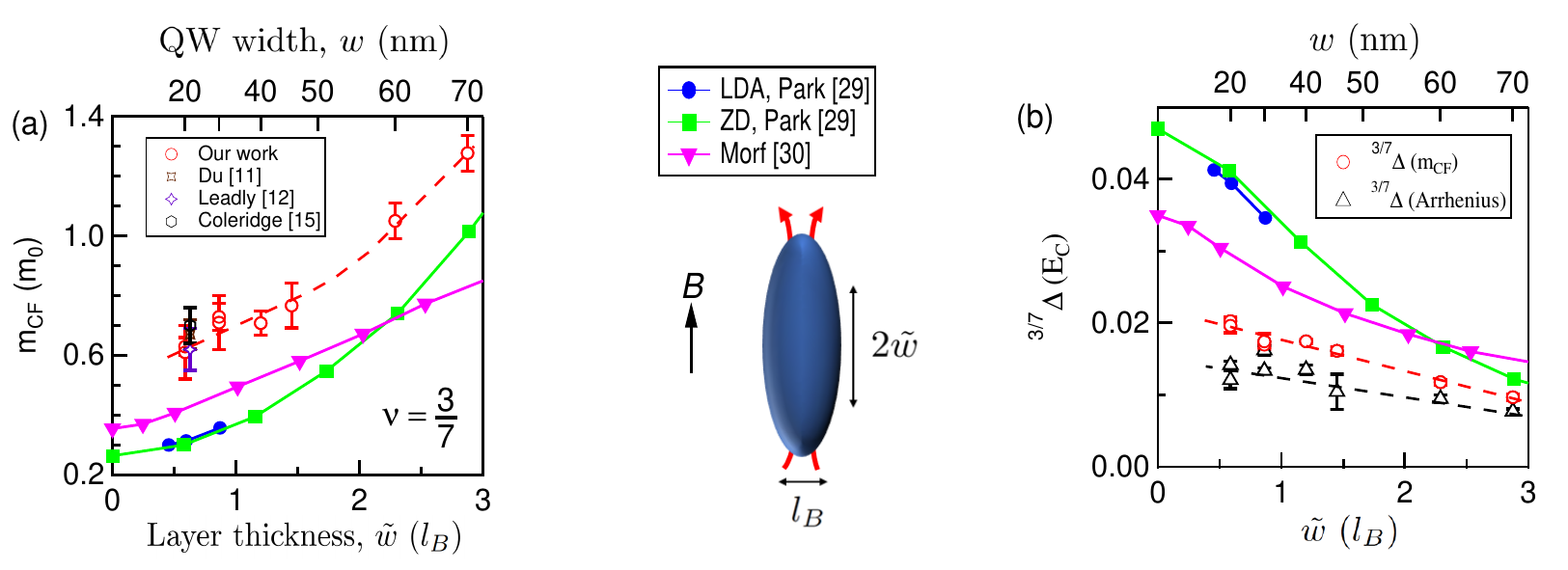, width=1\textwidth}
  \centering
  \caption{\label{Ip} 
(a) Red circles: Measured $m_{CF}$ vs. electron layer thickness $\tilde w$ for $\nu=3/7$. The data show that $m_{CF}$ increases for larger $\tilde w$. We have added three extra experimentally measured $m_{CF}$, as open brown, purple, and black symbols from Refs. \cite{Du.SSC.1994, Leadley.PRL.1994, Coleridge.PRB.1994}, respectively. The blue, green, and magenta closed symbols represent $m_{CF}$ derived from equating the theoretical energy gaps calculated in Refs. \cite{Park.Activation.1999} and \cite{Morf.PRB.2002} to the CF cyclotron gap. The calculations in Ref. \cite{Park.Activation.1999} are based on the local-density-approximation (LDA) and the Zhang-Das Sarma (ZD) potentials. (b) Open symbols: Energy gaps $^{3/7}\Delta$ vs. $\tilde w$ for $\nu=3/7$. The open red symbols are $^{3/7}\Delta$ obtained by converting our measured $m_{CF}$ to energy gaps, and the open black triangles are $^{3/7}\Delta$ measured from Arrhenius plots of $R_{xx}$ minimum vs. $1/T$ \cite{Kevin.PRL.2021}. The dashed lines are linear fits to the data points. The closed symbols are the theoretically-calculated $^{3/7}\Delta$ from Refs. \cite{Park.Activation.1999, Morf.PRB.2002}. The cartoon between the panels represents a two-flux CF with thickness 2$\tilde w$.
  }
  \label{fig:Ip}
\end{figure*}

Next, we focus on the second main contribution of our work, namely a comparison of our measured $m_{CF}$ to the results of available theoretical calculations that account for $\tilde w$. Park \textit{et al.} \cite{Park.Activation.1999} calculated the energy gaps ($\Delta$) of high-order FQHSs for 2DESs confined to $w=15$-, $20$-, and $30$-nm-wide GaAs QWs using two models: local-density-approximation (LDA), and the Zhang-Das Sarma (ZD) model interaction \cite{ZD}. The CF mass can be deduced by equating the theoretically-calculated $\Delta$ to the cyclotron energy, $\Delta=\hbar e B_{eff}/m_{CF}$. We plot $m_{CF}$ obtained as such using solid blue circles and green squares in Fig. 3(a) for the LDA and ZD interactions, respectively. Using exact diagonalization techniques, Morf \textit{et al.} \cite{Morf.PRB.2002} also calculated $\Delta$ as a function of $\tilde w$; we include $m_{CF}$ deduced from their $\Delta$ in Fig. 3(a) by solid magenta triangles. It is clear in Fig. 3(a) that there are major discrepancies between the measured and calculated $m_{CF}$ in the entire range of $\tilde w$. Since $m_{CF}$ should not be affected by disorder, this discrepancy is particularly surprising. 

For the sake of completeness, in Fig. 3(b) we show $^{3/7}\Delta$ calculated in Refs. \cite{Park.Activation.1999, Morf.PRB.2002} for $\nu=3/7$ vs. $\tilde w$ (closed symbols), as well as $^{3/7}\Delta$ we obtain by converting our measured $m_{CF}$ to energy gaps (open circles). As anticipated from Fig. 3(a), all $^{3/7}\Delta$ gaps decrease with increasing  $\tilde w$. In Fig. 3(b) we also show the $\nu=3/7$ energy gaps directly measured in our experiments from the Arrhenius plots of $R_{xx}$ minimum vs. $1/T$. The values for these directly measured $^{3/7}\Delta$ are the lowest and have a large scatter, likely reflecting the role of disorder in their determination. The $^{3/7}\Delta$ deduced from the experimentally measured $m_{CF}$ (open circles) are larger and have less scatter.  Nevertheless, they too fall below the calculated gaps (closed symbols), especially at small $\tilde w$. It is worth mentioning that the energy gaps measured (via Arrhenius plots) for the $\nu=1/3$ FQHS also exhibit the largest discrepancy with the theoretical values at smallest $\tilde w$ \cite{Kevin.PRL.2021}.

A few remarks are in order. First, it is worth emphasizing that the results presented here are complementary to the energy gap measurements (through Arrhenius plots) reported previously \cite{Du.PRL.1993, Pan.PRL.2020, Kevin.PRL.2021}. The dependence of the gap on layer thickness reported in Ref. \cite{Kevin.PRL.2021}, e.g., was primarily for the strongest FQHSs at $\nu=1/3$ and $2/3$. In our work here we focus on $m_{CF}$ for higher-order FQHSs. Indeed in our ultra-high-quality 2DESs, it is not possible to reliably determine $m_{CF}$ for the $1/3$ and $2/3$ FQHSs because the resistance oscillations on their flanks deviate from the expected sinusoidal form, thus rendering the application of Dingle analysis problematic \cite{Du.PRL.1994,Du.SSC.1994}. Second, in some samples, the measured $m_{CF}$ exhibit an apparent divergence for the highest-order FQHSs as $\nu=1/2$ is approached (see, e.g. Fig. S2(a) of the SM \cite{SM} for the 70-nm-wide QW sample). The origin of this anomalous divergence, which has also been reported before \cite{Manoharan.PRL.1994,Du.PRL.1994,Coleridge.PRB.1995}, remains a mystery. It is important to note that the filling factor ($\nu=3/7$) for which we present data in Fig. 3 is relatively far from $\nu=1/2$ to avoid the divergence complication (Fig. 2(a) and Fig. S2(a)). Third, in the SM \cite{SM}, we present data at three additional fillings, $\nu=4/9$, 5/11, and 4/7 which are also outside the divergence region. The conclusions described in the previous two paragraphs, namely the strong increase of $m_{CF}$ with increasing layer thickness and the discrepancy with the calculated values, also apply to data at these fillings.

We close by emphasizing that our reported $m_{CF}$ provide an ideal set of data for comparison with calculations, as they do not depend on disorder. The significant discrepancy between our measured $m_{CF}$ and available calculations is puzzling. It is tempting to attribute the discrepancy to the fact that the calculations did not include the role of LL mixing \cite{Footnote5}. Indeed, it is known that, at least for the principal FQHSs such as those at $\nu=1/3$ and $1/5$, such mixing would lower the FQHS energy gaps and thus raise $m_{CF}$ \cite{Yoshioka.JPSJ.1984,Yoshioka.SurfScience.1986,Melik-Alaverdian.PRB.1995, Kevin.PRL.2021, Sreejith.PRB.2017}. This could lead to a better agreement between the experimental data and calculations. However, it is worth mentioning that some recent calculations conclude that, surprisingly, including LL mixing might in fact \textit{lower} $m_{CF}$ \cite{Palacios.PRB.2021}. We hope that future calculations that more rigorously take into account the role of both electron layer thickness and LL mixing for high-order FQHSs would settle the discrepancies, and perhaps even explain the apparent divergence of $m_{CF}$ near $\nu=1/2$ \cite{Footnote4}.

\begin{acknowledgments}

We acknowledge support by the National Science Foundation (NSF) Grant No. DMR 2104771 for measurements. For sample characterization, we acknowledge support by the U.S. Department of Energy Basic Energy Sciences (Grant No. DEFG02-00-ER45841) and, for sample synthesis, NSF Grants No. ECCS 1906253 and MRSEC No. DMR 2011750, the Eric and Wendy Schmidt Transformative Technology Fund, and the Gordon and Betty Moore Foundation’s EPiQS Initiative (Grant No. GBMF9615 to L.N.P.). This research is funded in part by QuantEmX Travel Grants from the Institute for Complex Adaptive Matter. A portion of this work was performed at the National High Magnetic Field Laboratory (NHMFL), which is supported by National Science Foundation Cooperative Agreement No. DMR-1644779 and the state of Florida. We thank S. Hannahs, T. Murphy, A. Bangura, G. Jones, and E. Green at NHMFL for technical support. We also thank J. K. Jain and R. Winkler for illuminating discussions.
\end{acknowledgments}


\begin{thebibliography}{99}
\bibitem{Footnote5} In our samples, the LL mixing parameter defined as the ratio of Coulomb to cyclotron energies, is $0.83$ at $\nu=1/2$.
\bibitem{Footnote4} For future comparison of our measured $m_{CF}$ with new theoretical results, in the SM we also tabulate all our measured $m_{CF}$ at different fillings and QW widths.
\end{thebibliography}
\end{document}


\setcounter{page}{1}

\title[]{Supplementary Material for ``Composite Fermion Mass"}
\author{K. A. \surname{Villegas Rosales}}
\author{P. T. \surname{Madathil}}
\author{Y. J. \surname{Chung}}
\author{L. N. \surname{Pfeiffer}}
\author{K. W. \surname{West}}
\author{K. W. \surname{Baldwin}}
\author{M. \surname{Shayegan}}
\affiliation{Department of Electrical Engineering, Princeton University, Princeton, New Jersey 08544, USA}
\date{\today}

\maketitle   

\section{Data for the 30-nm-wide quantum well}

Figure S1 shows magnetoresistance data and Dingle analysis for a 30-nm-wide GaAs quantum well (QW) sample; sample $S_{30}$. We performed the Dingle analysis at fractional quantum Hall states (FQHSs) corresponding to the Landau level fillings $\nu=3/7$, $4/9$, $5/11$, $6/13$, $7/15$, and $8/17$. The Dingle analysis is carried out using the expression $\Delta R/R_{0}=4 \times exp(-\pi/\omega_{CF}\tau _{q})\times \xi/sinh(\xi)$ \cite{Dingle.1952,Shoenberg.1984,Du.PRL.1994,Du.SSC.1994,Manoharan.PRL.1994,Leadley.PRL.1994}. The factor $\xi/sinh(\xi)$ represents the $T$-induced damping, where $\xi= 2\pi^2k_{B}T/\hbar\omega_{CF}$, and $\tau _{q}$ is the CF quantum lifetime. Other relevant parameters are defined as $\Delta R=(R_{max}-R_{min})$ and $R_{0}=(R_{max}+R_{min})/2$, where $R_{max}$ and $R_{min}$ are defined Fig. 1(b) of the main text.

The plots for the Dingle analysis at several fractional fillings are displayed in Figs. S1(b-g). Note that the Dingle fits, denoted by red curves in these figures, are done so that they fit the experimental data points well at high temperatures.  At lower temperatures (below about 0.2 K), the experimental points fall below the fitted curves for $\nu=3/7$, $4/9$, $5/11$, and $6/13$. This is because at these fillings, the resistances of the FQHSs saturate at zero and constrain $R_{min}$. The saturation limits the growth of $\Delta R/R_0$ at low temperatures, and leads to a deviation of the experimental data points relative to what is predicted by the Dingle expression (red curves).

\section{Data for the 70-nm-wide quantum well}

Figure S2 summarizes our data for the 70-nm-wide GaAs QW sample; sample $S_{70}$. In Fig. S2(a) the measured $m_{CF}$ are shown by open circles as a function of $B^{1/2}$ at several $\nu$ where FQHSs are observed. The $m_{CF}$ data points follow an approximately linear trend as a function of $B^{1/2}$. Note that, qualitatively similar to the $S_{30}$ data, $m_{CF}$ for very high-order FQHSs ($p>7$) deviate from the dashed line, pointing to a possible divergence of $m_{CF}$ as $\nu=1/2$ is approached. The dashed line in Fig. S2(a) is a linear fit to the data, excluding the points associated with the divergence. 

In Fig. S2(b) the black triangles present $^{\nu}\Delta$ vs. (\textit{e}$^{2}$/$4\pi\epsilon_{0}\epsilon$\textit{l}$_{B}$)/$(2p+1)$ measured for the same sample, for $\nu$ up to $8/17$ and $8/15$. We determined $^{\nu}\Delta$ from the standard procedure of making Arrhenius plots of $R_{xx}$ minimum as a function of $1/T$ and fitting the data to $R_{xx}\propto exp(-^{\nu}\Delta/2T)$. In Fig. S2(b), we also plot the values for $^{\nu}\Delta$, by red open circles, deduced from our measured $m_{CF}$ (from Fig. S2(a)) and using the expression $^{\nu}\Delta=\hbar \omega_{CF} = \hbar e B_{eff}/m_{CF}$. The black lines are fits to the triangles, and they have negative intercepts with the $y$-axis. The average absolute value of these negative intercepts provides an estimate of the disorder energy $\Gamma$, by which the activation energy gaps (triangles) are reduced. The red lines are linear fits to the red open circles. The intercept of the red lines with the $y$-axis is effectively zero, consistent with the assumption that the deduced $m_{CF}$ are insensitive to disorder.

\section{Data for the 60-nm-wide quantum well}

Figure S3 shows $m_{CF}$ and energy gap data for the sample with a $60$-nm-thick GaAs QW; sample $S_{60}$.

\section{Composite fermion quantum lifetime $\tau_{q}$}

There are two factors in the Dingle expression, $exp(-\pi/\omega_{CF}\tau _{q})$ and $\xi/sinh(\xi)$ that determine the amplitude on the magnetoresistance oscillations. In these expressions, $\tau _{q}$ is the CF quantum lifetime, $\hbar \omega_{CF} = \hbar e B_{eff}/m_{CF}$ is the CF cyclotron energy, and $\xi= 2\pi^2k_{B}T/\hbar\omega_{CF}$. It is generally assumed that these two factors are disentangled, at least for certain form of disorder-broadening of the Landau levels \cite{Shoenberg.1984}. In this scenario, an analysis of the temperature dependence of the amplitude of an oscillation near a fixed magnetic field, based on the $\xi/sinh(\xi)$ term, would yield a value for the effective mass. This is what we have done in Figs. 1(c) and 1(d) of the main text and Figs. S1(b-g). 

It is also possible to deduce the CF quantum lifetime $\tau_{q}$ from the so-called ``Dingle plots" which describe the magnetic field dependence of the magnetoresistance oscillations (at a fixed temperature) according to $exp(-\pi/\omega_{CF}\tau _{q})$. To include data taken at different temperatures, it customary to plot $(\Delta R/R_{0})(\xi/sinh(\xi))$ vs. $(\pi m_{0}/e)(m_{CF}/|B_{eff}|)$ \cite{Dingle.1952,Shoenberg.1984,Du.PRL.1994,Du.SSC.1994,Manoharan.PRL.1994,Leadley.PRL.1994}. (The parameter $\Delta R/R_{0}$ is defined as before; see Fig. 1(b)). In Fig. S4 we show such a Dingle plot for the $30$-nm-thick QW, sample $S_{30}$. We observe an approximately linear behavior; the red line is a fit to the data points and yields a value of $\simeq18$ for $\tau_{q}$. From similar analyses of data, we obtain $\tau_{q} \simeq38$ and $29$ ps for the 60-, and 70-nm-thick GaAs QWs, respectively. Using $\Gamma_{\tau_{q}}=\hbar/\tau_{q}$, we can deduce a corresponding disorder parameter. For $S_{30}$, $S_{60}$, and $S_{70}$, $\Gamma_{\tau_{q}}$ are $\simeq 0.42$, $0.21$, and $0.27$ K, respectively. These values are about 3 to 4 times smaller than $\Gamma$ deduced from the FQHS energy gap measurements, namely, from the intercepts of black lines in Figs. 2(b), S2(b), and S3(b). This discrepancy provides further evidence that the intercepts of the black lines in Figs. 2(b), S2(b) and S3(b) may not be directly or solely related to disorder, as we also discussed in the main text. We would like to emphasize that the very good fits observed in Fig. 1(c) and Fig. S4 attest to the appropriateness of the Dingle expression in describing the experimental data, and the disentanglement of the two terms of the expression in determining the CF Landau levels' temperature broadening, which yields $m_{CF}$ (Fig. 1(c)), and disorder broadening, which gives $\tau _{q}$ (Fig. S4).




\section{Composite Fermion masses, and Energy gaps derived from measured masses at different fillings}

Figure S5 displays $m_{CF}$ and $^{4/7}\Delta$ vs. $\tilde w$. In Fig. S5(b), we have inverted our measured $m_{CF}$ to determine $\Delta$ and show $\Delta$ by open red circles. The black symbols are measured from Arrhenius plots of $R_{xx}$ vs $1/T$. We compare these to the theoretically-available energy gaps from Refs. \cite{Park.Activation.1999, Morf.PRB.2002}.

Figures S6 and S7 provide similar $m_{CF}$ and $\Delta$ data for $\nu=4/9$ and $5/11$, respectively.


\clearpage

\begin{figure*}[h!]
  \centering
    \psfig{file=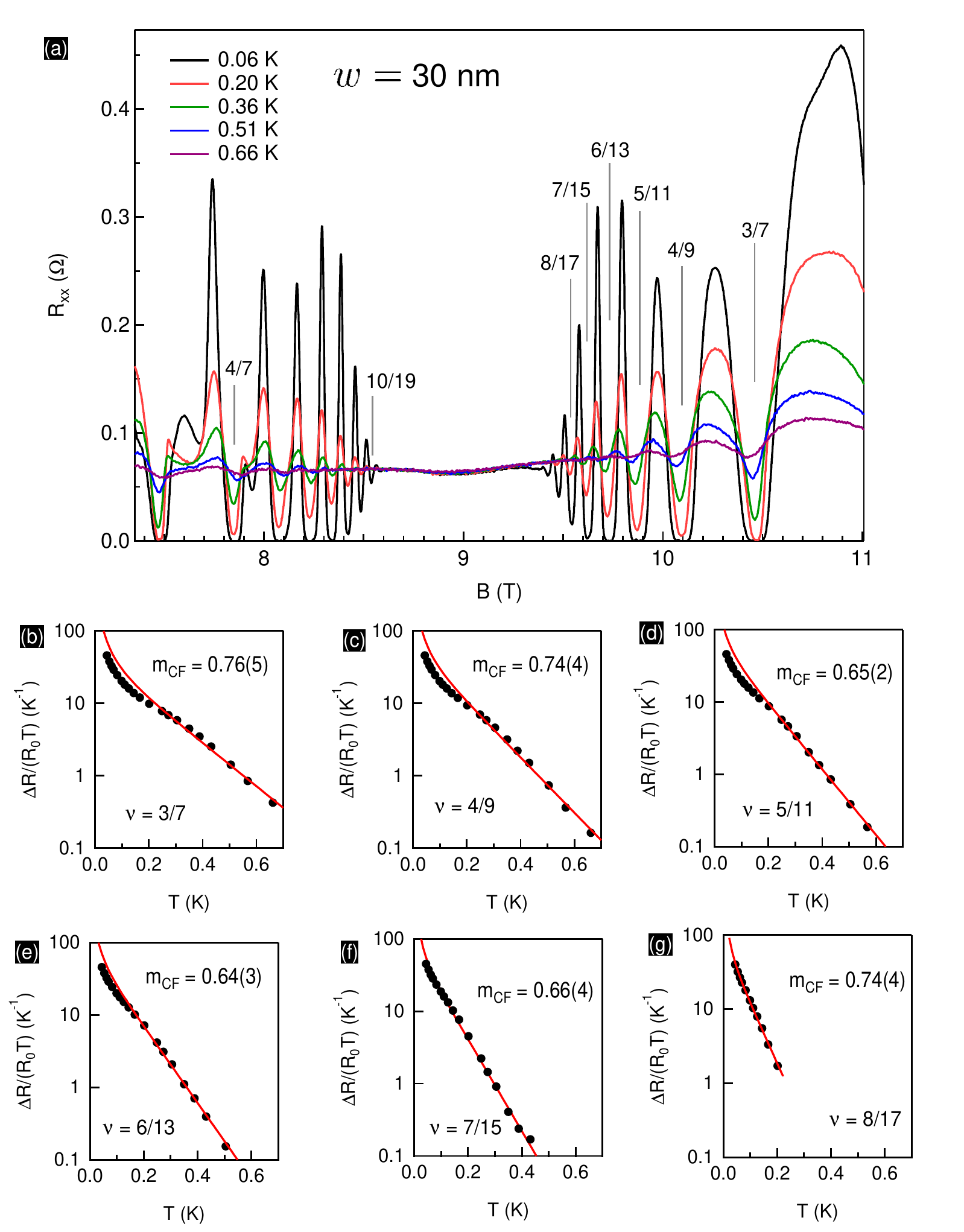, width=0.9\textwidth}
  \centering
  \caption{\label{transport} 
   (a) Longitudinal magnetoresistance $R_{xx}$ vs. $B$ for $S_{30}$ at different temperatures. (b-g) Dingle analysis for $\nu=3/7$, $4/9$, $5/11$, $6/13$, $7/15$, and $8/17$. The red curves are fits to the experimental data points at high temperatures.
  }
  \label{fig:transport}
\end{figure*}

\clearpage

\begin{figure*}[h!]
  \centering
    \psfig{file=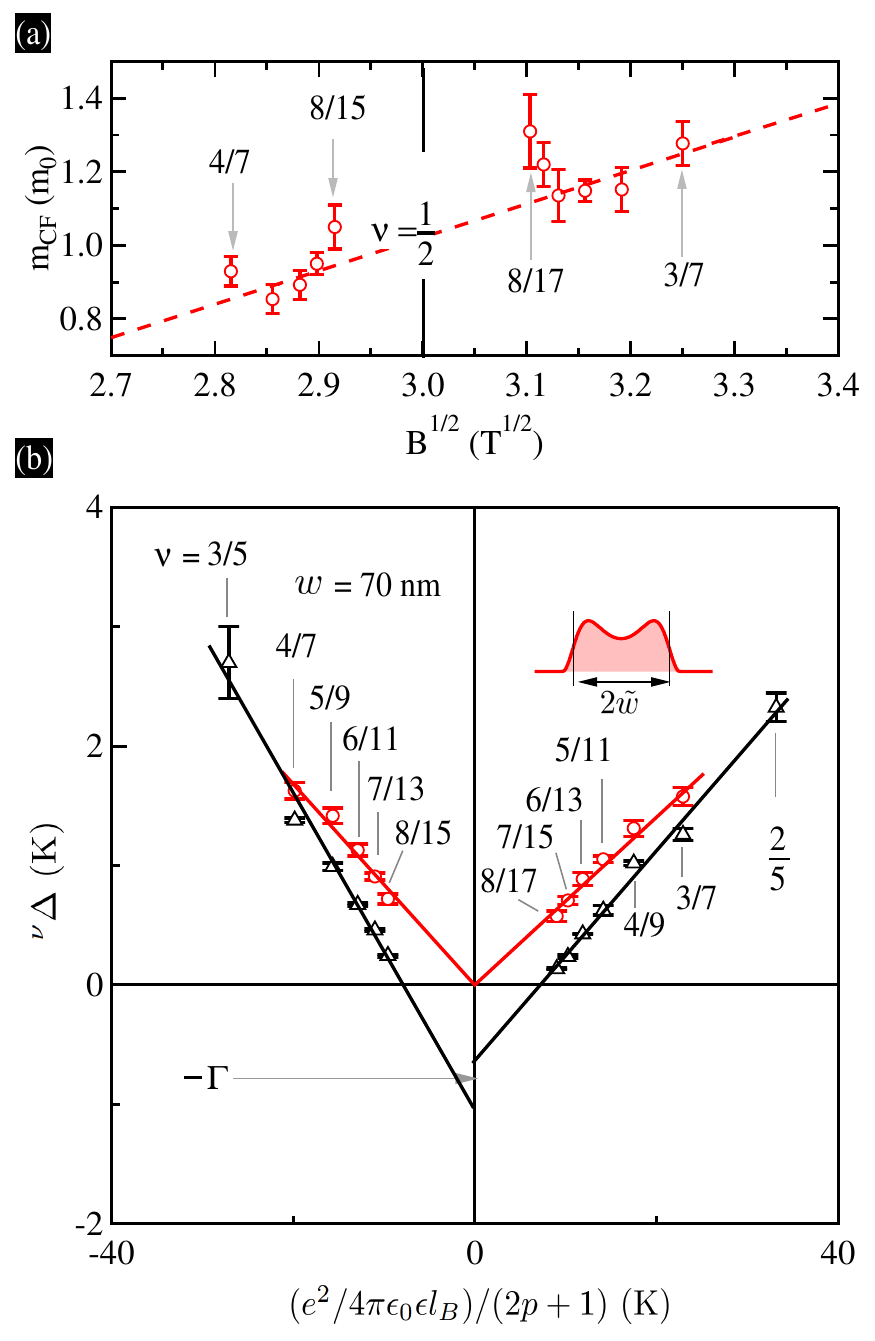, width=0.65\textwidth}
  \centering
  \caption{\label{transport} 
   (a) Composite fermion effective mass $m_{CF}$ vs. $B^{1/2}$ for a 2DES confined to a $70$-nm-wide GaAs QW. The open circles are $m_{CF}$ determined at FQHS fillings. (b) Black symbols are FQHS energy gaps $^{\nu}\Delta$, measured from Arrhenius plots of $R_{xx}$ vs. $1/T$. They are shown as a function of $(e^{2}/4\pi\epsilon_{0}\epsilon l_{B})/(2p+1)$. The black lines are linear fits to the black triangles, and their negative intercepts yield an estimate for the disorder paramater $\Gamma=(1.1\pm0.2)$ K. The open red circles are $^{\nu}\Delta$ deduced from the cyclotron energy of CFs, $\hbar \omega_{CF}=\hbar e B_{eff}/m_{CF}$, where we use the values of $m_{CF}$ shown in (a). The red lines are fits to the open red circles and their intercept is $(-0.04\pm0.10)$ K.
  }
  \label{fig:transport}
\end{figure*}

\clearpage

\begin{figure}[h!]
  \centering
    \psfig{file=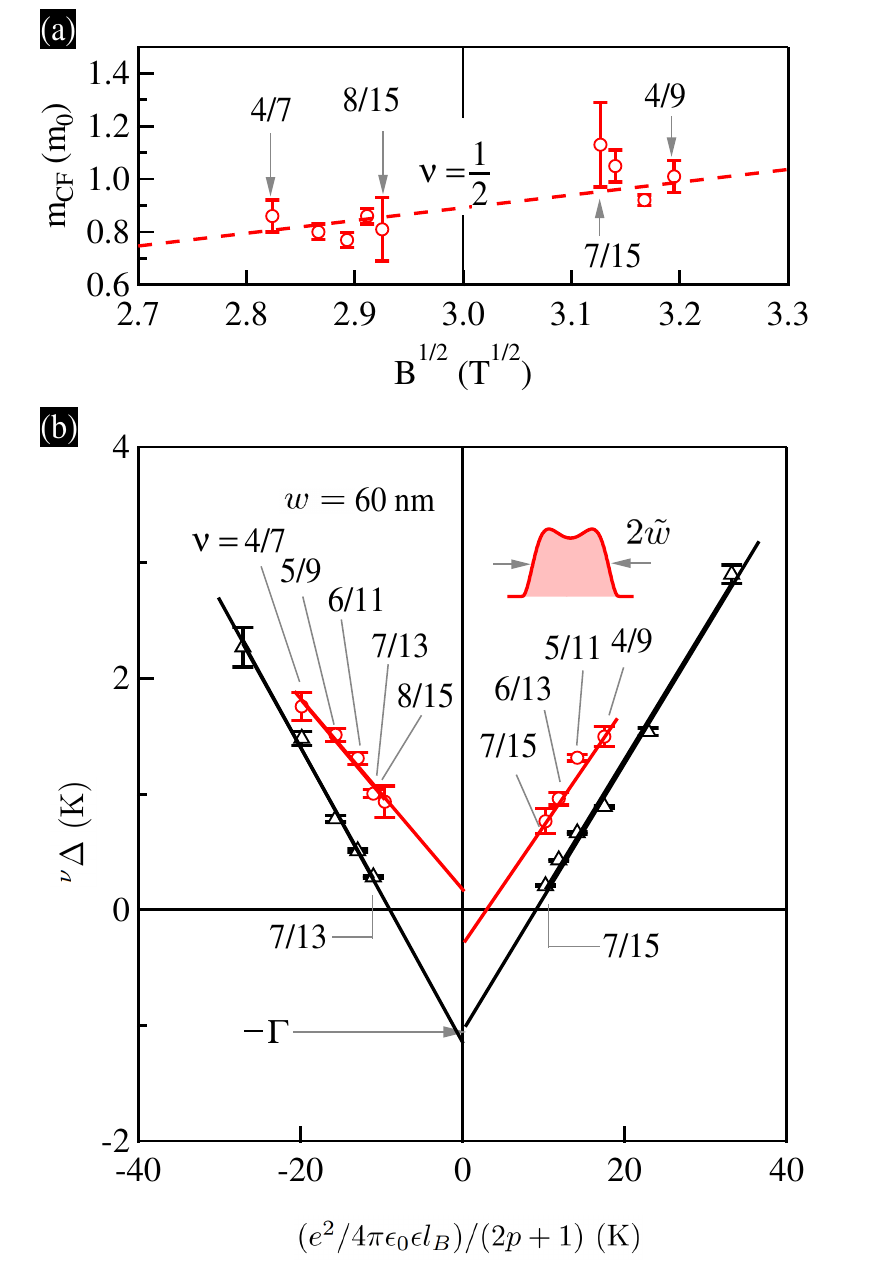, width=0.7\textwidth}
  \centering
  \caption{\label{transport} 
   (a) Composite fermion effective mass $m_{CF}$ vs. $B^{1/2}$ for a 2DES confined to a $60$-nm-wide GaAs QW. (b) Black symbols are FQHS energy gaps $^{\nu}\Delta$, measured from Arrhenius plots of $R_{xx}$ vs. $1/T$. They are shown as a function of $(e^{2}/4\pi\epsilon_{0}\epsilon l_{B})/(2p+1)$. The black lines are linear fits to the black triangles, and their negative intercepts yield an estimate for the disorder paramater $\Gamma=(1.1\pm0.2)$ K. The open red circles are $^{\nu}\Delta$ deduced from the cyclotron energy of CFs, $\hbar \omega_{CF}=\hbar e B_{eff}/m_{CF}$, where we use the values of $m_{CF}$ shown in (a). The red lines are fits to the open red circles and their intercept is $(-0.03\pm0.22)$ K.
  }
  \label{fig:transport}
\end{figure}

\clearpage

\begin{figure}[h!]
  \centering
    \psfig{file=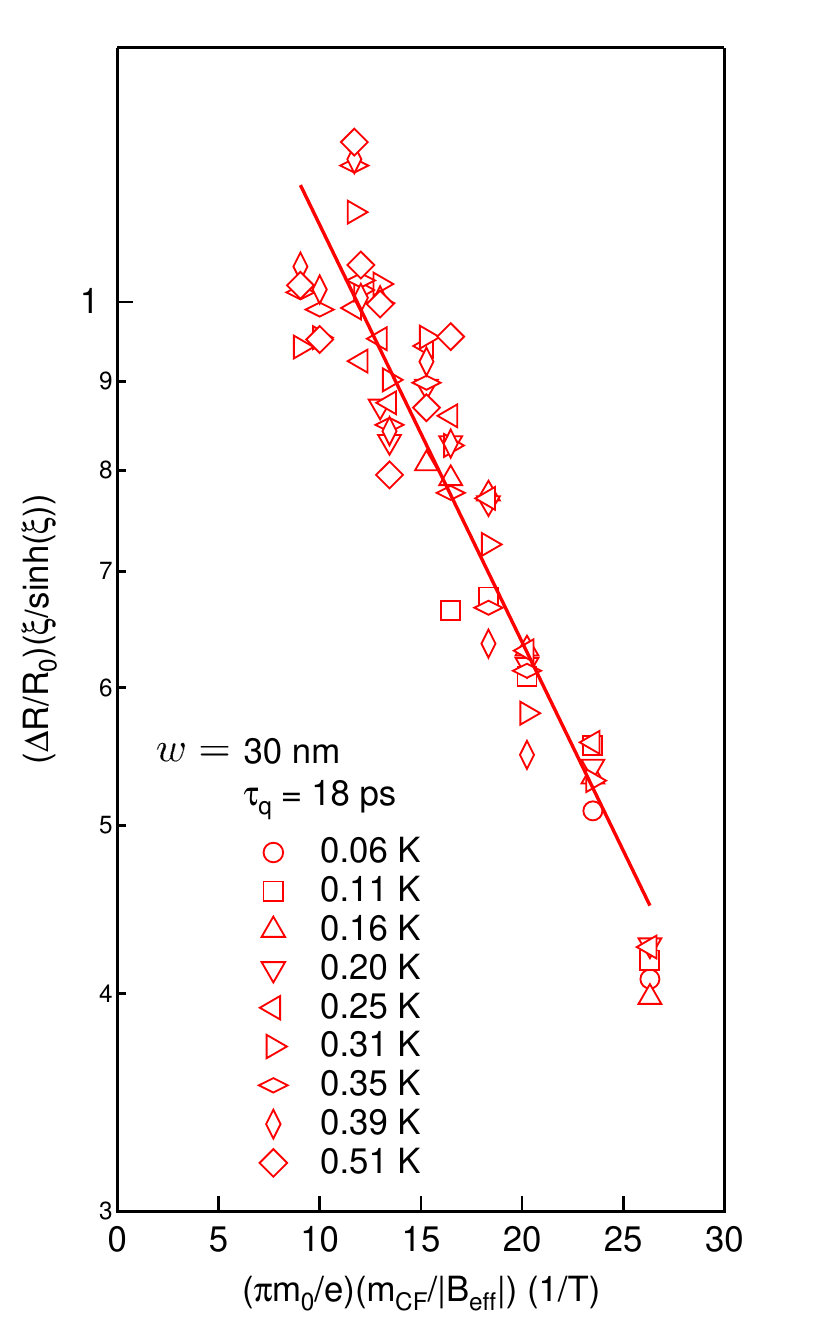, width=0.7\textwidth}
  \centering
  \caption{\label{transport} 
   Dingle plot for the 30-nm-thick GaAs QW sample. The open and closed symbols represent data taken at the FQHS fillings. The deduced quantum lifetime, deduced from fitting the Dingle expression to the data (red line), is $\tau_{q}\simeq18$ ps.  The corresponding $\Gamma_{\tau_{q}}=\hbar/\tau_{q}\simeq 0.42$ K.
  }
  \label{fig:transport}
\end{figure}

\clearpage


\clearpage


\clearpage

\clearpage

\begin{figure}[h!]
  \centering
    \psfig{file=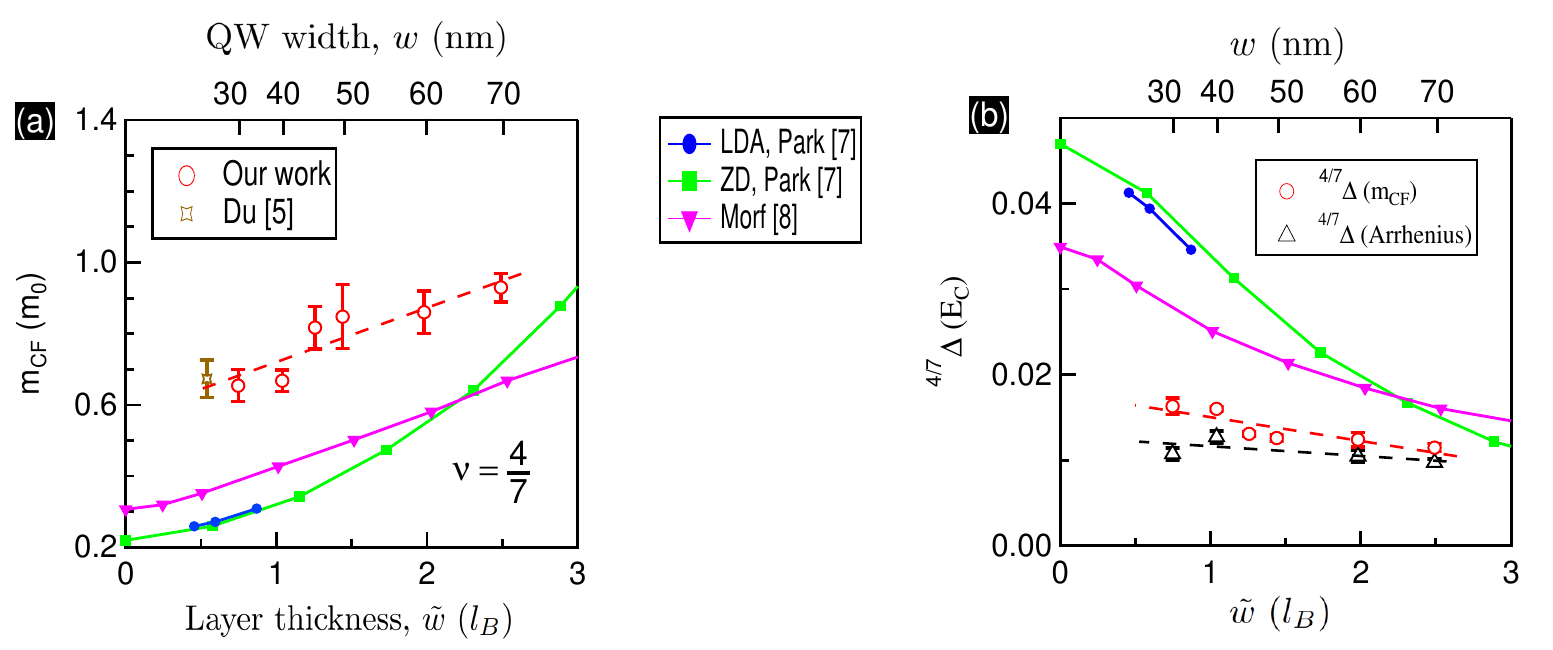, width=1\textwidth}
  \centering
  \caption{\label{transport} 
(a) Red circles: Measured CF effective mass $m_{CF}$ vs. electron layer thickness $\tilde w$ for $\nu=4/7$. The data show that $m_{CF}$ increases for larger $\tilde w$. We have added one extra experimentally measured $m_{CF}$, as an open brown symbol from Du \textit{et al.} \cite{Du.SSC.1994}. The blue, green, and magenta symbols and curves represent $m_{CF}$ derived from equating the theoretically-calculated energy gaps to the CF cyclotron gap; blue and green points are from Ref. \cite{Park.Activation.1999} and magenta points are from \cite{Morf.PRB.2002}. The calculations in Ref. \cite{Park.Activation.1999} are based on the local-density-approximation (LDA) and the Zhang-Das Sarma (ZD) potentials. (b) Open symbols: Energy gaps $^{4/7}\Delta$ vs. $\tilde w$ for $\nu=4/7$. The open red symbols are $^{4/7}\Delta$ obtained by converting our measured $m_{CF}$ to energy gaps, and the open black symbols are $^{4/7}\Delta$ measured from Arrhenius plots of $R_{xx}$ minimum vs. $1/T$. The dashed lines are linear fits to the open symbols. The closed symbols are the theoretically-calculated $^{4/7}\Delta$; blue and green are from Ref. \cite{Park.Activation.1999} and magenta are from Ref. \cite{Morf.PRB.2002}. 
  }
  \label{fig:transport}
\end{figure}

\begin{figure}[h!]
  \centering
    \psfig{file=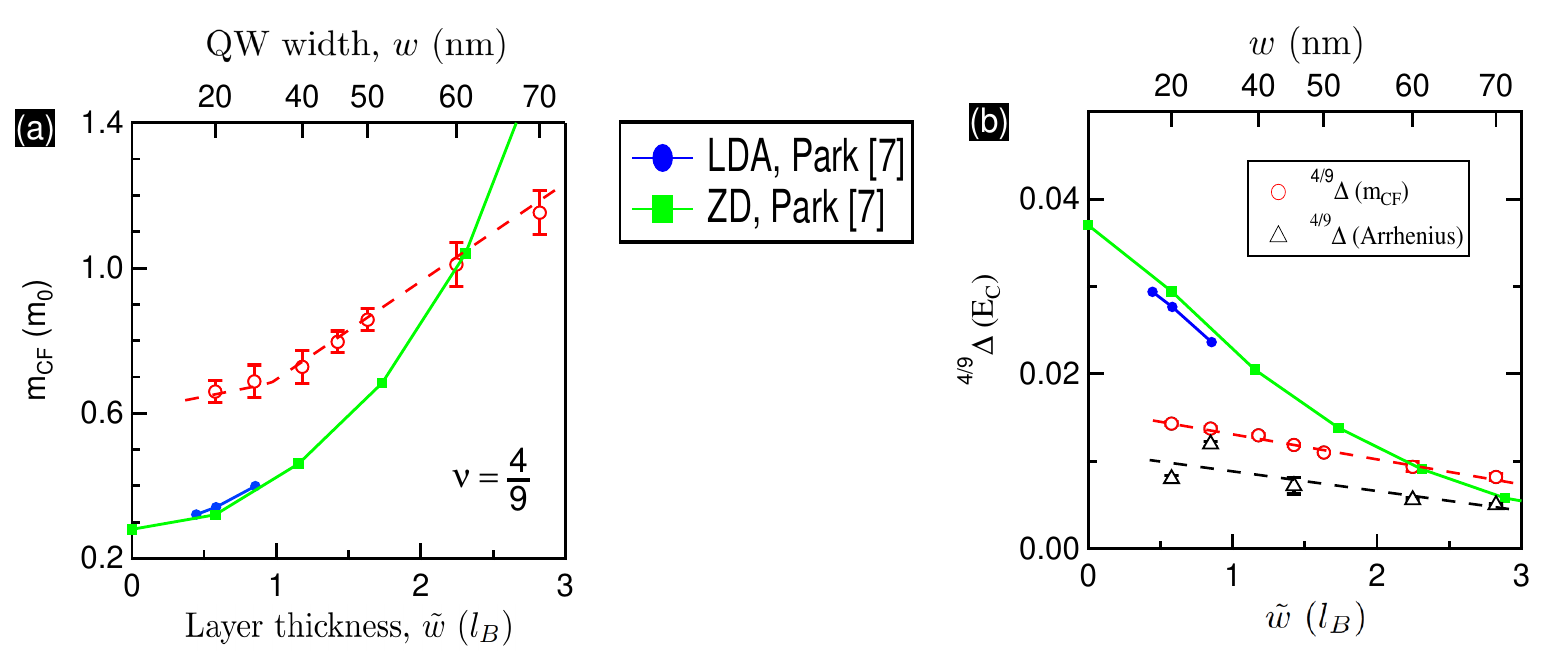, width=1\textwidth}
  \centering
  \caption{\label{transport} 
(a) Black circles: Measured CF effective mass $m_{CF}$ vs. electron layer thickness $\tilde w$ for $\nu=4/9$. The data shows that $m_{CF}$ increases for larger $\tilde w$. The blue and green symbols and curves represent $m_{CF}$ derived from equating the theoretically-calculated energy gaps to the CF cyclotron gap; blue and green points are from Ref. \cite{Park.Activation.1999}. For completeness, and easier comparison with the results of past and future calculations, in Table I we list the values of $m_{CF}$ measured in our experiments. The calculations in Ref. \cite{Park.Activation.1999} are based on the local-density-approximation (LDA) and the Zhang-Das Sarma (ZD) potentials. (b) Open symbols: Energy gaps $^{4/9}\Delta$ vs. $\tilde w$ for $\nu=4/9$. The red symbols are $^{4/9}\Delta$ obtained by converting our measured $m_{CF}$ to energy gaps, and the black symbols are $^{4/9}\Delta$ measured from Arrhenius plots of $R_{xx}$ minimum vs. $1/T$. The dashed lines are linear fits to the open symbols. The closed symbols are the theoretically-calculated $^{4/9}\Delta$; blue and green are from Ref. \cite{Park.Activation.1999} and red are from Ref. \cite{Morf.PRB.2002}. 
  }
  \label{fig:transport}
\end{figure}

\begin{figure}[h!]
  \centering
    \psfig{file=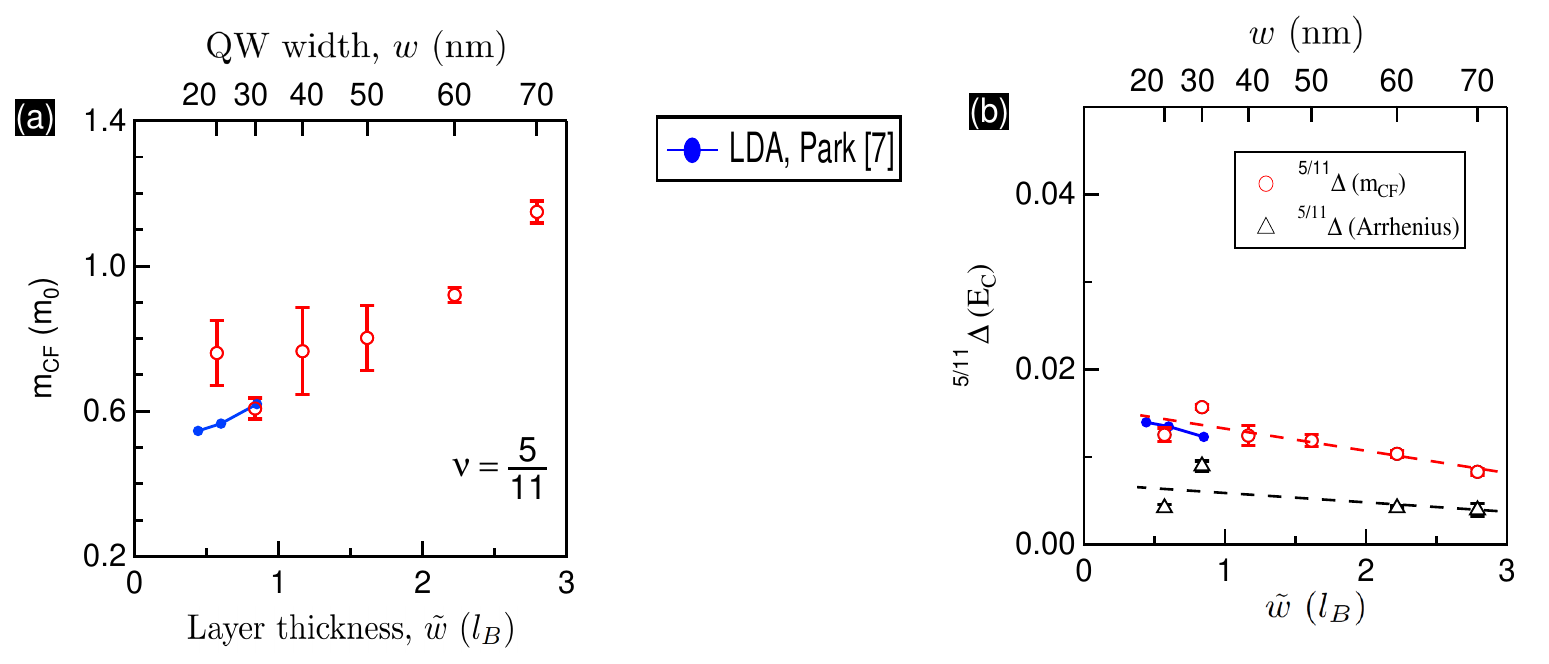, width=1\textwidth}
  \centering
  \caption{\label{transport} 
(a) Black circles: Measured CF effective mass $m_{CF}$ vs. electron layer thickness $\tilde w$ for $\nu=5/11$. The data shows that $m_{CF}$ increases for larger $\tilde w$. The blue closed circles and curve represent $m_{CF}$ derived from equating the theoretically-calculated energy gaps to the CF cyclotron gap; blue points are from Ref. \cite{Park.Activation.1999}. The calculations in Ref. \cite{Park.Activation.1999} are based on the local-density-approximation (LDA). (b) Open symbols: Energy gaps $^{5/11}\Delta$ vs. $\tilde w$ for $\nu=5/11$. The red symbols are $^{5/11}\Delta$ obtained by converting our measured $m_{CF}$ to energy gaps, and the black symbols are $^{5/11}\Delta$ measured from Arrhenius plots of $R_{xx}$ minimum vs. $1/T$. The dashed lines are linear fits to the open symbols. The blue closed circles are the theoretically-calculated $^{5/11}\Delta$ are from Ref. \cite{Park.Activation.1999}. 
  }
  \label{fig:transport}
\end{figure}

\clearpage

\section{Table of measured $m_{CF}$ values at different $\nu$ and quantum well widths}

For completeness, and for easier comparison with the results of past and future calculations, in Table I we list the values of $m_{CF}$ measured in our experiments. The numbers in parentheses give the error in the last digit; e.g., $0.71(4)$ means $0.71 \pm 0.04$.

\begin{table}[!h]
\begin{center}
\begin{tabular}{ |c||c|c|c|c|c|c|c|c|c| } 
 \hline
 $\nu$ & $S_{20}^{1}$ & $S_{20}^{2}$ & $S_{30}^{1}$ & $S_{30}^{2}$ & $S_{40}$ & $S_{45}$ &  $S_{50}$ & $S_{60}$ & $S_{70}$ \\
 \hline
 $2/5$  &   & $0.48(9)$ &  & $0.73(5)$ &  & & & &\\
 \hline
 $3/7$  & $0.63(6)$ & $0.61(9)$ & $0.76(5)$ & $0.72(2)$ & $0.71(4)$ & $0.77(8)$ & & & $1.28(6)$ \\
 \hline
 $4/9$  & $0.66(3)$ & & $0.74(4)$ & & $0.73(4)$ & $0.80(3)$ & $0.86(3)$ & $0.86(3)$ & $1.15(6)$ \\
 \hline
 $5/11$  & $0.76(9)$ & & $0.65(2)$ & & $0.76(14)$ & & & $0.80(9)$ & $1.15(3)$ \\
 \hline
 $6/13$  & & & $0.64(3)$ & & & & & $0.85(9)$ & $1.13(7)$ \\
 \hline
 $7/15$  & & & $0.66(4)$ & & & & & $0.13(16)$ & $1.22(6)$ \\
 \hline
 $8/17$  & & & $0.74(4)$ & & & & & & $1.3(1)$ \\
 \hline
 \hline
 $8/15$  & & & $0.69(2)$ & & & & & $0.8(1)$ & $1.05(6)$ \\
 \hline
 $7/13$  & & & $0.67(2)$ & & & & & $0.86(3)$ & $0.95(3)$ \\
 \hline
 $6/11$  & & & $0.61(3)$ & & & & & $0.77(3)$ & $0.89(4)$ \\
 \hline
 $5/9$  & & &  & & $0.72(1)$ & & & $0.80(3)$ & $0.85(4)$ \\
 \hline
 $4/7$  & & & & $0.66(4)$ & $0.67(4)$ & $0.82(4)$ & $0.85(9)$ & $0.86(6)$ & $0.93(4)$ \\
\hline
\end{tabular}
\caption{Measured composite fermion masses.}
\label{table:1}
\end{center}
\end{table}